\documentclass[twocolumn]{aastex631}
\usepackage{amsmath}
\pdfoutput=1

\usepackage{cuted}
\usepackage{xcolor}

\begin{document}

\title{Galaxy-galaxy lensing data: $f(T)$ gravity challenges General 
Relativity}

\author[0000-0003-0194-0697]{Qingqing Wang}
\affiliation{Department of Astronomy, School of Physical Sciences, University of Science and Technology of China, Hefei 230026, China}
\affiliation{CAS Key Laboratory for Researches in Galaxies and Cosmology, School of Astronomy and Space Science, University of Science and Technology of China, Hefei, Anhui 230026, China}
\affiliation{Deep Space Exploration Laboratory, Hefei 230088, China}

\author[0000-0002-5450-0209]{Xin Ren}
\affiliation{Department of Physics, Tokyo Institute of Technology, Tokyo 152-8551, Japan}
\affiliation{Department of Astronomy, School of Physical Sciences, University of Science and Technology of China, Hefei 230026, China}
\affiliation{CAS Key Laboratory for Researches in Galaxies and Cosmology, School of Astronomy and Space Science, University of Science and Technology of China, Hefei, Anhui 230026, China}
\affiliation{Deep Space Exploration Laboratory, Hefei 230088, China}

\author[0000-0002-3784-8684]{Bo Wang} 
\affiliation{Department of Astronomy, School of Physical Sciences, University of Science and Technology of China, Hefei 230026, China}
\affiliation{CAS Key Laboratory for Researches in Galaxies and Cosmology, School of Astronomy and Space Science, University of Science and Technology of China, Hefei, Anhui 230026, China}
\affiliation{Deep Space Exploration Laboratory, Hefei 230088, China}

\author[0000-0003-0706-8465]{Yi-Fu Cai}
\affiliation{Department of Astronomy, School of Physical Sciences, University of Science and Technology of China, Hefei 230026, China}
\affiliation{CAS Key Laboratory for Researches in Galaxies and Cosmology, School of Astronomy and Space Science, University of Science and Technology of China, Hefei, Anhui 230026, China}
\affiliation{Deep Space Exploration Laboratory, Hefei 230088, China}

\author[0000-0003-1297-6142]{Wentao Luo}
\affiliation{Department of Astronomy, School of Physical Sciences, University of Science and Technology of China, Hefei 230026, China}
\affiliation{CAS Key Laboratory for Researches in Galaxies and Cosmology, School of Astronomy and Space Science, University of Science and Technology of China, Hefei, Anhui 230026, China}
\affiliation{Deep Space Exploration Laboratory, Hefei 230088, China}

\author[0000-0003-1500-0874]{Emmanuel N. Saridakis}
\affiliation{National Observatory of Athens, Lofos Nymfon, 11852 Athens, Greece}
\affiliation{Department of Astronomy, School of Physical Sciences, University of Science and Technology of China, Hefei 230026, China}
\affiliation{Departamento de Matem\'{a}ticas, Universidad Cat\'{o}lica del Norte, Avda. Angamos 0610, Casilla 1280 Antofagasta, Chile}
\footnote{Corresponding authors: yifucai@ustc.edu.cn; wtluo@ustc.edu.cn; msaridak@noa.gr}

\begin{abstract}
We use galaxy-galaxy lensing data to test General Relativity and $f(T)$ gravity at galaxies scales. We consider an exact spherically symmetric solution of $f(T)$ theory which is obtained from an approximate quadratic correction, and thus it is expected to hold for every realistic deviation from General Relativity. Quantifying the deviation by a single parameter $Q$, and following the post-Newtonian approximation, we obtain the corresponding deviation in the gravitational potential, shear component,  and effective excess surface density profile. We used five stellar mass samples and divided them into blue and red to test the model dependence on galaxy color, and we modeled the excess surface density (ESD) profiles using the Navarro-Frenk-White (NFW) profiles. Based on the group catalog from the Sloan Digital Sky Survey Data Release 7 (SDSS DR7) we finally extract $Q=-2.138^{+0.952}_{-0.516}\times 10^{-5}\,$Mpc$^{-2}$ at $1\sigma$ confidence. This result indicates that $f(T)$ corrections on top of General Relativity are favored. Finally, we apply information criteria, such as the AIC and BIC ones, and although the dependence of $f(T)$ gravity on the off-center effect implies that its optimality needs to be carefully studied, our analysis shows that $f(T)$ gravity is more efficient in fitting the data comparing to General Relativity and $\Lambda$CDM paradigm, and thus it offers a challenge to the latter. 
\end{abstract}

\section{Introduction} \label{sec:intro}

The concordance model of cosmology, namely $\Lambda$CDM paradigm, incorporates General Relativity as the underlying gravitational theory, the standard model of particles, cold dark matter (CDM) and the cosmological constant $\Lambda$, and it has been well verified by various observational datasets, such as  Cosmic Microwave Background (CMB) \citep{Planck:2015, Planck:2018}, Baryon Acoustic Oscillations (BAO) \citep{SDSS:2005, BOSS:2016}, Type Ia supernovae\citep{SupernovaSearchTeam:2004, SNLS:2005},  galaxy formation and evolution theory \citep{Davis:1985}, as well as weak lensing observation \citep{Heymans:2012, Shi:2017qpr}. However, recent observations have revealed possible tensions \citep{Bullock:2017, Perivolaropoulos:2021jda}, such as the Hubble tension between early-time measurements under $\Lambda$CDM and late-time local  distance-ladder measurements \citep{Wong:2019, Abdalla:2022yfr}, and the $\sigma_{8}$ tension between the relative clustering level found in CMB experiments and the late time large-scale structure observations \citep{DiValentino:2020vvd, Yan:2019gbw}. While, the non-renormalizability of General Relativity and the difficulties in bringing it closer to a quantum description is a potential disadvantage for the theory \citep{Addazi:2021xuf}. Hence, a significant amount of research has been devoted in the construction of various gravitational modifications, aiming to alleviate or solve the above issues \citep{Capozziello:2011et, CANTATA:2021ktz}.

Modified Newtonian dynamics (MOND) theory was one of the firs that gained 
recognition as a possible scheme for extragalactic dynamics phenomenology 
\citep{Bekenstein:2004ne}. Nevertheless, one can proceed in modifying General Relativity, resulting to  modified and extended theories of gravity. To obtain this, one starts from the Einstein-Hilbert action and 
incorporate extra terms, obtaining $f(R)$ gravity \citep{Starobinsky:1980te, Capozziello:2002rd}, $f(G)$ gravity \citep{Nojiri:2005jg}, Weyl gravity \citep{Mannheim:1988dj} and Lovelock gravity \citep{Lovelock:1971yv}. However, one could start from the equivalent formulation of gravity in terms of torsion 
\citep{Maluf:2013gaa}, 
and follow similar procedures, resulting to $f(T)$ gravity \citep{Cai:2015emx, Krssak:2018ywd, Bahamonde:2021gfp}, $f(T,T_G)$ gravity
\citep{Kofinas:2014owa}, $f(T,B)$ gravity \citep{Bahamonde:2015zma} etc. 
These classes of theories have been shown to present very interesting 
phenomenology \citep{Cai:2018rzd, Ren:2021tfi, Ren:2022aeo, Hu:2023juh, Hu:2023xcf}. Furthermore, one may take into account non-metricity, and construct symmetric teleparallel gravity and $f(Q)$ gravity \citep{BeltranJimenez:2017tkd, Anagnostopoulos:2021ydo}.

Every theory of gravity, including General Relativity, ought to pass various tests, using a variety of observational data \citep{Berti:2015}, 
from the expansion of the universe to the formation of large-scale structures. Tests of gravity on small scales usually study the consistency of its cosmological feasibility alongside Solar System experiments, and in order to achieve this it is necessary to first quantify possible deviations from General Relativity and then use the data to constrain them 
\citep{Will:2014kxa, Chan:2022, Chiba:2006}. As it is known, at the Solar System level, General Relativity is always inside the obtained parametric contours for the various modified gravity parameters (for the case of $f(T)$ gravity see \citep{Iorio:2012cm, Bahamonde:2020}).

On the other hand, with the development of large-scale galactic surveys, weak gravitational lensing has become increasingly important in delineating matter distribution, leading it to become a powerful tool in constraining modified gravity  \citep{Bacon:2000sy, Luo:2017zbc,Cai:2023ite}. In \citep{Chen:2019} the authors performed for the first time a novel test on possible deviations form General Relativity using galaxy-galaxy weak gravitational lensing. They used $f(T)$ framework to quantify these deviations and then used the deflection angle at non-cosmological scales \citep{Ruggiero:2016} to approximately calculate the lensing potential and the effective surface mass density, and thus extract the upper bound on the deviation parameter with the weak lensing data from SDSS DR7. Additionally, the perturbative spherically symmetric solution of the covariant formula of $f(T)$ theory was extracted within $T+\alpha T^{2}$ deviation from GR in \citep{Ren:2021}, where the deflection angle and the difference in position and magnification in the lensing frame were calculated.

In this work we desire to employ galaxy-galaxy weak gravitational lensing in order to extract more accurate constraints on possible deviations from General Relativity. Using $f(T)$ gravitational theories in order to quantify the deviation, interestingly enough we find that the quadratic correction on top of GR is favored. The plan of the manuscript is the following: 
In Section \ref{fttheory} we briefly review $f(T)$ gravity and we present 
the spherically symmetric solutions. Then, in Section \ref{wlensing} we calculate the corresponding gravitational potential and the weak lensing shear signal, and we derive a correction term related to the negative quadratic radius. In Section \ref{data} we introduce the group catalog \citep{Yang:2007} and the shear catalog from the SDSS DR7 \citep{SDSS:2008} in order to extract observational constraints. Hence, in Section \ref{results} we fit the ESD and we provide the estimation results for the model parameters, alongside the application of AIC and BIC information criteria. Finally, Section \ref{conclution} is dedicated to conclusions and outlook.

\section{Specifically symmetric solutions in $f(T)$ ravity}\label{fttheory} 

In this section we use $f(T)$ gravity in order to quantify possible deviations form General Relativity, and we extract the corresponding 
corrections on specifically symmetric solutions. 

In the framework of teleparallel gravity one uses the tetrad field  
$h^{A}{}_{\mu}$, related to the metric through $g_{\mu \nu 
}=h^{A}{}_{\mu}h^{D}{}_{\nu}\eta_{AD}$, where $\eta_{AD}=\text{diag}(1, -1, -1, 
-1)$. Concerning the connection, one uses the teleparallel one, namely \citep{Cai:2015emx}
\begin{equation}
 \Gamma^{\rho}{}_{\nu \mu} = h_{A}{}^{\rho}\partial h^{A}{}_{\nu}+h_{A}{}^{\rho}\omega^{A}{}_{D\mu}h^{D}{}_{\nu} ~,
\end{equation}
where $\omega^{A}{}_{D\mu}$ represents a flat metric-compatible spin
connection, and therefore the torsion tensor is  
\begin{align}
    {T}^{\rho}{}_{\mu \nu}& \equiv {\Gamma}^{\rho}{}_{\nu 
\mu}-{\Gamma}^{\rho}{}_{\mu \nu} \nonumber
\\ &=h_{A}{}^{\rho}(\partial_{\mu} 
h^{A}{}_{ 
\nu}-\partial_{\nu} h^{A}{}_{\mu}+{\omega}^{A}{}_{D \mu} 
h^{D}{}_{\nu}-{\omega}^{A}{}_{D \nu} h^{D}{}_{ \mu}) ~. 
\end{align}
Furthermore, the torsion scalar is defined as
\begin{equation}
T = S_{\rho }{}^{\mu \nu}  T^{\rho }{}_{\mu \nu} ~, 
\end{equation}
where ${S_{\rho}}^{\mu \nu} \equiv \frac{1}{2}\left({{K}^{\mu 
\nu}}_{\rho}+\delta_{\rho}^{\mu} {T^{\alpha \nu}}_{\alpha}-\delta_{\rho}^{\nu} 
{T^{\alpha \mu}}_{\alpha}\right)$ is the super-potential, 
with the  contortion tensor being
${K}^{\rho}{}_{\mu \nu} \equiv 
\frac{1}{2}\left(T_{\mu}{}^{\rho}{}_{\nu}+T_{\nu}{}^{\rho}{}_{\mu}-{T^{\rho}}_{
\mu \nu}\right)$.
The action of $f(T)$ gravity is
\begin{equation}
    S=\int d^{4} x \frac{h}{16 \pi G}f(T),
\end{equation}
where $h=det(h^{A}{}_{\mu})$ is the tetrad determinant and $G$ the gravitational constant. Finally, performing variation in terms of the tetrad, gives rise to the field equations of $f(T)$ gravity as
\begin{align}
     & h(f_{T T} h_{A}{}^{\rho} {S}_{\rho}{}^{\mu \nu} \partial_{\nu} {T}  
-f_{T}h_{A}{}^{\rho} {T}^{\lambda}{}_{\nu \rho} {S}_{\lambda}{}^{\nu \mu}  
\nonumber \\ & +f_{T} \omega^{D}_{\ A \nu} h_{D}{}^{\rho} {S}_{\rho}{}^{\nu \mu} 
+\frac{1}{2} f h_{A}{}^{\mu} )+f_{T} \partial_{\nu}\left(h h_{A}{}^{\rho} 
{S}_{\rho}{}^{\mu \nu}\right)  \nonumber \\
 & = 4\pi G h h_{A}{}^{\rho} \overset{(m)}{T}_{\rho}{}^{\mu} ~.
\end{align}

In the following we will work in the Weitzenb$\ddot{\text{o}}$ck gauge, in 
which the spin connection vanishes \citep{Cai:2015emx}. Solving the anti-symmetric part of the field equations with vanishing spin connection one can obtain the general spherically symmetric tetrad. There are two different branches of tetrad solution ans$\ddot{\text{a}}$tze in this 
approach. The first branch, corresponding to real tetrad, is a familiar case and has been well studied \citep{Tamanini:2012hg, DeBenedictis:2016aze, Bahamonde:2019zea, Ruggiero:2015oka, Ren:2021}. The second branch is a complex tetrad, namely \citep{Bahamonde:2021}  
\begin{align} 
\label{eq:tetrad}
  &h^{A}{}_{\mu}= \\ \nonumber
&\small{\left (\begin{array}{cccc}
0 &iB(r) &0 &0 \\
iA(r)sin \theta cos \psi &0 &-\chi r sin \psi & -r\chi sin  \theta cos\theta cos 
\psi  \\
iA(r)sin \theta sin \psi &0 & \chi r cos \psi & -r\chi sin  \theta cos\theta sin 
\psi \\
iA(r)cos \theta &0 &0 &\chi sin^{2} \theta \\
\end{array}\right) } ~,
\end{align}
and the corresponding metric is  
\begin{equation}
\label{eq:metric}
ds^{2}=A(r)^{2}dt^{2}-B(r)^{2}dr^{2}-r^{2}d\Omega^{2} ~,
\end{equation}
where $\chi= \pm 1$. We mention that although the tetrad is complex, all 
physical quantities, such as the metric, the torsion scalar and the boundary term are real and independent of the sign of $\chi$. 

One can see that the general field equations accept the exact solution for the metric function $A(r)$: 
\begin{equation}
A(r)^{2}=1-2\frac{M}{r}+\frac{Q}{r^{2}}  ~,
\end{equation}
if 
\begin{equation}
f(T) = \frac{4(2\pm P)}{(QT+2\pm P)(8-2QT\pm 4P)} ~,
\label{eq:model1}
\end{equation}
where $P=\sqrt{Q^{2}T^{2}-2QT+4}$, in which case the  torsion scalar is 
$T=(4r^{2}-6Q)/(r^{4}-2Qr^{2})$. Note that for $Q\ll 1$ this model can be 
expanded as 
\begin{equation}
 f(T)= T-\frac{1}{8}QT^{2} +{\cal{O}}(T^3).
\label{eq:model2}
\end{equation}
$Q$ is the single parameter that quantifies deviations from General Relativity, and for $Q=0$ the latter is recovered.

In summary, the physically meaningful metric solution of model \eqref{eq:model2} in complex tetrad \eqref{eq:tetrad} is written as  \citep{Bahamonde:2021}
 {\small{
\begin{equation}
\label{eq:metricsolution}
ds^{2}=\left(1\!-\!\frac{2M}{r}\!+\! 
\frac{Q}{r^{2}}\right)\!dt^{2}-\left(\frac{2Mr\!-\!Q\!-\!r^{2}}{2Q-r^{2}}
\right)^{ -1 } \!dr^ { 2 } -r^ { 2 } d 
\Omega^{2 } ~,
\end{equation}}}
which is an exact spherically symmetric solution in $f(T)$ gravity. We stress that $Q$ is a parameter of the theory, completely independent of the electromagnetic charge, and the Schwarzschild solution is recovered for $Q = 0$.

\section{Weak lensing }\label{wlensing}

In this section we present the weak lensing machinery. The propagation of a photon in the universe is determined by the spacetime properties, and the local matter inhomogeneity leads to a deflection of its trajectory, resulting to the lensing effect \citep{Bartelmann:1999}. Hence, the spacetime properties, and thus the underlying theory of gravity, leave 
imprints in the lensing signal contains, which can then be used as a test of the theory of gravity itself.

We consider that light from a distant source is affected by the gravitational field in the foreground during its propagation. Since the effective speed of light is reduced in a gravitational field, it will delay relatively to vacuum propagation. At the same time, photons are also   deflected when they pass through a gravitational field. Thus, the difference between $f(T)$ gravity and General Relativity can be reflected in the deflection angle $\hat{\alpha}$ in static spherically symmetric spacetime.

We assume that the whole lensing system lies in the asymptotically flat 
spacetime regime. The distance of the light ray  of closest approach $r_0$, as well as the impact parameter $b$, both lie outside the gravitational 
radius, and the deflection angle for metric \eqref{eq:metric} is given 
as \citep{Keeton:2005}
\begin{align}
    \hat{\alpha}(r_{0}) &=2\int^{\infty}_{r_{0}}\left |\frac{d\phi}{dr}\right  
|dr-\pi \notag\\
    &=2\int^{\infty}_{r_{0}}\frac{1}{r^2}\sqrt{\frac{A^{2}B^{2}}{1/b^2-A^{2}/r^2}}dr - \pi   ~,
\end{align}
where the impact parameter $b=\sqrt{\frac{r^2_0}{A^{2}(r_0)}}$ is given by the 
vertical distance of the asymptotic tangent of the ray trajectory from the 
center of the lens to the observer with respect to the inertial observer.
Imposing the identification $m = \frac{GM}{c^2}$, the bending angle can then be expressed as a series expansion in the single quantity $\frac{m}{b}$ as \citep{Keeton:2005}
\begin{equation}
\hat{\alpha}(b)=A_{1}\left(\frac{m}{b}\right)+A_{2}\left(\frac{m}{b}\right)^2+A_
{3}\left(\frac{m}{b}\right)^3+\mathcal{O}\left(\frac{m}{b}\right)^4 ~.
\end{equation}
For the spherically symmetric metric solution \eqref{eq:metricsolution} 
the deflection angle can be calculated under the parametrized-post-Newtonian (PPN) formalism. Following \citep{Keeton:2005}, we calculate the coefficients of the first three orders as
\begin{equation}
 A_{1}=4, 
 A_{2}=\frac{5}{4}\pi \left(3-\frac{Q}{m^{2}}\right), 
 A_{3}=\frac{64}{3}\left(2-\frac{Q}{m^{2}}\right)  ~.
\end{equation}

The effective lensing potential is defined as \citep{Narayan:1996ba}:
\begin{equation}
 \Psi (\vec{\theta})=\frac{2 D_{ds}}{c^2D_s D_d}\int \Phi(D_d \theta,z) dz ~,
\end{equation}
where $D_d$, $D_{ds}$, and $D_s$ denote the angular diameter distances between observer and lens, lens and source, and observer and source, respectively. The gravitational potential of the lens can be derived from the relationship between deflection angle and effective lensing potential
under the second-order approximation like: 
\begin{equation}
 \hat{\alpha} = \frac{2}{c^2}\int \nabla_{D_d \theta} \Phi(D_d \theta,z) dz ~,
\end{equation}
and thus we acquire the three-dimensional gravitational potential:
\begin{equation}
 \Phi = -\frac{GM}{r} - \frac{15(GM)^{2}}{8r^{2}c^2}+\frac{5Qc^2}{8r^{2}} ~.
\end{equation}
Here the first term is the Newtonian potential, the second one is the 
contribution from General Relativity, and the third term is the modification from $f(T)$ gravity. Note that the second term was neglected in \citep{Chen:2019}, hence its incorporation in the present work will lead to significantly more accurate results.

Then, the shear tensor can be calculated by the linear combination of the 
second partial derivatives of the potential, namely
\begin{align}
 &\gamma_1 \equiv \frac12\!\left(  \frac{\partial^2 \Psi}{\partial\theta_1^2} - \frac{\partial^2 \Psi}{\partial\theta_2^2} \right) \equiv \gamma\cos2\phi \notag 
\\ &=
     -\frac{2D_{ds}D_d}{c^2 D_s} \!
\left(\!\frac{2\Delta\Sigma(R)}{\pi}\!-\!\frac{15\pi 
Qc^2}{16R^{3}}\!+\!\frac{45\pi (GM)^{2}}{16R^{3}c^2}\right)\!\cos2\phi
\end{align}
\begin{align}
    &\gamma_2 \equiv  \frac{\partial^2 \Psi}{\partial\theta_1\partial\theta_2} = 
\frac{\partial^2 \Psi}{\partial\theta_2\partial\theta_1} \equiv \gamma\sin2\phi 
\notag \\ &=  -\frac{2D_{ds}D_d}{c^2 
D_s}\left(\!\frac{2\Delta\Sigma(R)}{\pi}\!-\!\frac{15\pi 
Qc^2}{16R^{3}}\!+\!\frac{45\pi 
(GM)^{2}}{16R^{3}c^2}\right)\!\sin2\phi,
\end{align}
where $\Delta\Sigma(R)$ is the ESD under Newtonian approximation. Hence, in order to distinguish we define an effective ESD as:
\begin{align}
    \Delta \Sigma(R)_{eff} &=\Sigma(\leq R)-\Sigma(R)=\gamma\Sigma_{crit}  
\notag \\
    &=\Delta \Sigma(R)-\frac{15 Qc^2}{32GR^{3}}+\frac{45 GM^{2}}{32R^{3}c^2}  ~,
\end{align}
with $\Sigma_{crit}=\frac{c^{2}}{4\pi G}\frac{D_{s}}{D_{ds}D_d}$ the critical surface mass density. As we can see, the shear components $\gamma_1$ and $\gamma_2$ introduce anisotropy (or astigmatism) into the lens mapping, and the quantity $\gamma = (\gamma_{1}^{2} + \gamma_{2}^{2})^{1/2}$ describes the magnitude of the shear. Since our metric is chosen from the black hole solution, the result is the point source approximation (see \citep{Turyshev:2023huw} for cases under the extended source in Solar gravitational lens circumstance).  

Since we are dealing with possible deviations form General Relativity, our 
predictions should be confronted with observational data form galaxies, 
large-scale structure, and dark matter halo density profiles. The matter distribution of the lens can be described by the modification of the 
NFW density profile \citep{Mandelbaum:2004}. In 
particular, the NFW profile consists of two parameters, the characteristic mass scale $M_{200}$ and the halo concentration $c_h$, which is defined as the ratio between the virial radius of a halo and its characteristic scale radius. 

In the following section we will use the ESD calculated under the NFW model in order to describe the Newtonian term, and we will adopt the oncentration-mass relation to reduce the degrees of freedom.

\section{Data confrontation}\label{data} 

We have now all the machinery required for a confrontation with observational data. In order to test the excess surface density caused by $f(T)$ gravity, we use a weak lensing catalog provided by SDSS DR7. Concerning the lens samples, we use the ones from the spectroscopic group catalog \citep{Yang:2007} based on SDSS DR7, which applies a halo-based group finding algorithm \citep{Yang:2006zf}. The algorithm first assumes that each galaxy is a potential candidate group of galaxies, and then it calculates the total luminosity of each galaxy. The various galaxies were identified by selecting information such as distance and redshift that meet the criteria, and the sample was selected from individual galaxy systems to attenuate the bias caused by other surrounding structures, finally resulting to a sample size of 400,608. 

The galaxy's stellar mass is calculated by the stellar mass-to-light ratio 
from \citep{Bell:2003}. Specifically, we use the data given by \citep{Luo:2020} which are divided into five stellar mass bins. In order to test the influence produced by $f(T)$ gravity on galaxy formation time, we further subdivide our sample into blue galaxies and red galaxies, based on a cut of a color-magnitude plane from \citep{Yang:2007}, given by:
\begin{align}
    ^{0.1}(g-r) = 1.022-0.0652x-0.0031x^2 ~.
\end{align}
In this expression $g$ and $r$ are the different band magnitudes, $x = 
^{0.1}M_{r}-5\textrm{log}h+23.0$, and $^{0.1}M_{r}-5\textrm{log}h$ is the 
absolute magnitude of the galaxy after $K$ correction, while the superscript describes evolution correction at redshift $z=0.1$ \citep{SDSS:2004dnq}. 

We use the shape catalog created by \citep{Luo:2016}, based on SDSS DR7 imaging data for the source. SDSS imaging data of DR7 contains about 230 million different luminosity objects, covers about 8,423 square degrees of LEGACY sky, including the $u$, $g$, $r$, $i$, and $z$ five-band photometry. 
Thus, in \citep{Luo:2016} the authors built an image processing pipeline to 
correct systematic errors introduced mainly by the point spread function (PSF). This pipeline processes the SDSS DR7 $r$ band imaging data, which can be used to generate the background galaxy catalog containing the shape information of each galaxy. The final shape catalog that we use contains the position, shape, shape error, and photometric redshift information of about 40 million galaxies, based on \citep{Csabai:2007}.

The ESD $\Delta\Sigma(R)$ that we use as the shear implies that it is measured by the weighted mean of source galaxy shapes, namely
\begin{align}
 \Delta\Sigma(R)= \frac{1}{2\Bar{R}}\frac{\Sigma (w_{i}e_{t}(R)\Sigma_{cls})}{\Sigma w_{i}} ~,
\end{align}
where $w$ is the weight for each source galaxy, calculated by the shape noise $\sigma_{shape}$ and the noise from sky $\sigma_{sky}$
\begin{align}
    w=\frac{1}{(\sigma_{shape}^{2}+\sigma_{sky}^{2})} ~.
\end{align}

\section{results}\label{results} 

\begin{figure*}
\centering
\includegraphics[width=15cm]{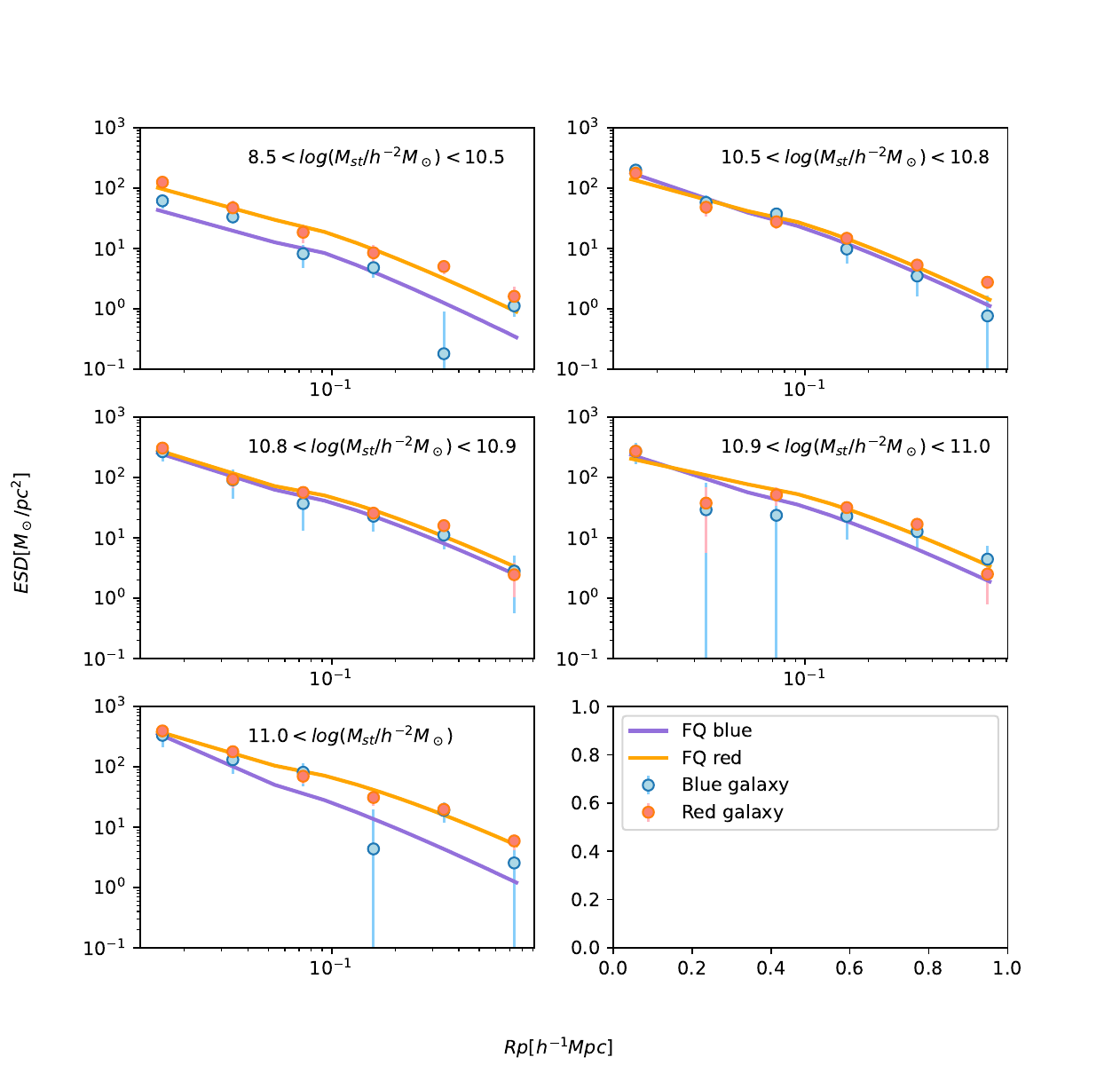}
\caption{{\it{Best fits to the ESD profile. The horizontal axis is the projected distance away from the lens galaxy. The blue and orange points with error bars represent the weak lensing data for blue and red galaxies, while the purple and orange lines are the fit curves. Here we choose $R_{sig}=0.01$.}}}
\label{fig:ksquarefitting}
\end{figure*}

In order to detect the distribution  mechanism of excess surface density 
calculated under modified gravity, we use the signal of the weak gravitational lensing to constrain the model parameter $Q$. Firstly, we consider a modified NFW model with General Relativity and $f(T)$ corrections. Secondly, since the lens sample is selected from an individual galaxy system, and the contribution of the modified model parameters only played a significant role on small scales, we perform the analysis without taking into account the contribution of the two halo terms.
The distribution of  galaxies determines the location of the center of the lens gravitational potential, and the selected central galaxy is not necessarily at the center of the gravitational potential, which is called the off-center effect. The off-center effect can influence the estimation of the mass distribution because it adds significant system uncertainty. Therefore, we consider a slightly off-center effect of about $R_{sig}=0.01h^{-1}Mpc$ for a conservative estimation \citep{Luo:2017zbc}.

We proceed by using ten catalogs of the five mass intervals of the red 
and blue galaxies in order to test ESD calculated under $f(T)$ gravity, and then we obtain the best-fit results of the model by using the Chi-square test. The $\chi^{2}$ is calculated as 
\begin{align}
    \chi^{2} =  
((\Delta\Sigma_{data}-\Delta\Sigma_{eff})^{T}C^{-1}(\Delta\Sigma_{data}
-\Delta\Sigma_{eff})) ~,
\end{align}
where $C^{-1}$ is the inverse covariance matrix.

We mention that the  measurements  of the small stellar mass bin have very high signal-to-noise ratios. Moreover, the correction term of $f(T)$ gravity contributes significantly to the small mass end, hence it plays an important role in reducing $\chi^{2}$. Furthermore, we take into account the dependence of the weak lensing data and the $f(T)$ correction on the color characteristics of the galaxies in the fitting process. As can be seen from Fig.~\ref{fig:ksquarefitting}, since there are fewer blue galaxies in the large mass bin, the ESD profile carries less information. This results in a decrease in the limit level of the model parameters in the last two bins of the blue galaxy, while the red galaxy has a better overall signal quality.

Since the signal-to-noise ratio of red galaxies is significantly  better than that of blue galaxies, we mix them up to extract information from five 
different stellar mass bins to constrain the model parameters. Finally, as a very conservative estimation, we adopt the concentration-mass relationship proposed by \citep{Neto:2007vq} as a Gaussian before eliminating degeneracies with other parameters, and we use the MCMC method to obtain the final parameter space of $Q$.

\begin{figure*}
\label{fig:Parameterconstrain}
\centering
\includegraphics[width=15cm]{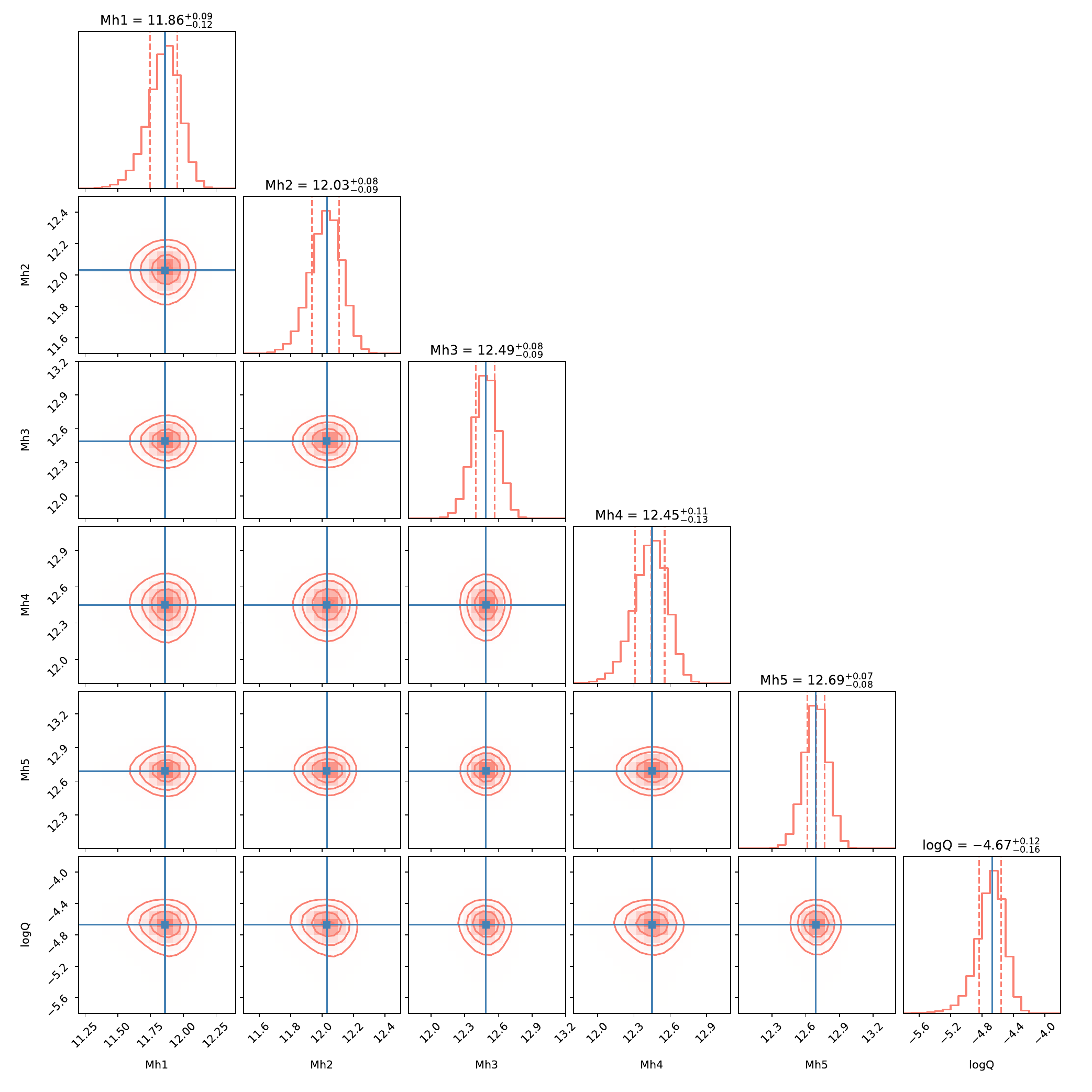}
\caption{Constraints at 68.0\% , 95.0\% and 99.0\% confidence level, for 
halo mass and deviation parameter. Here we choose $R_{sig}=0.01$.}
\label{fig:constrain}
\end{figure*}

In summary, after performing the above steps, we finally obtain the 
estimation for the $Q$ interval within 1$\sigma$ confidence level as
\begin{align}
Q=-2.138^{+0.952}_{-0.516}\times 10^{-5} Mpc^{-2} ~.
\end{align}
Additionally, in Fig.~\ref{fig:constrain} we present the two-dimensional posterior of the fitting, as well as the constraints for various quantities.

As we observe, the $f(T)$-gravity parameter $Q$ that quantifies the deviation from General Relativity, and interestingly enough the value zero is excluded at 1$\sigma$ confidence level. This suggests that $f(T)$ corrections on top of General Relativity are favoured, at least at the galaxy scales, where galaxy-galaxy lensing data are sensitive to the gravitational potential. This is one of the main result of the this work. 

In order to compare the fitting quality and the model performance with that of standard $\Lambda$CDM concordance scenario, we apply the Akaike Information Criterion (AIC) and the Bayesian Information Criterion (BIC), 
\citep{liddle2007information}. The AIC criterion provides an estimator of the Kullback-Leibler information, it exhibits the property of asymptotic 
unbiasedness, and is defined as \citep{Akaike:1974vps}:
\begin{align}
 AIC \equiv -2ln\mathcal{L}_{max}+2p_{tot} ~, 
\end{align}
where $\mathcal{L}_{max}$ represents the maximum likelihood of the model, and $p_{tot}$ represents the total number of free parameters. Additionally, the BIC criterion provides an estimator of the Bayesian evidence, and is defined as \citep{Kass:1995loi}
\begin{align}
 BIC \equiv -2ln\mathcal{L}_{max}+p_{tot}ln(N_{tot}) ~,
\end{align}
where $N_{tot}$ is the number of samples, while the other parameters are the same as in AIC. According to Jeffreys classification \citep{Kass:1995loi}, if $\Delta\text{IC}\leq 2$ then the scenario is statistically compatible with the best-fit model, if $2<\Delta\text{IC}<6$ then we have a moderate tension between the two scenarios, while for $\Delta\text{IC}\geq 10$ we obtain significant tension. Hence, we apply these criteria to draw a comparison between the NFW model in $\Lambda$CDM and the modified NFW model in $f(T)$ gravity. 

\begin{table} 
\begin{center}
\begin{tabular}{c|c|c|c|c}
\hline\hline
Model  & $BIC$ & $AIC$ &$\Delta BIC$ & $\Delta AIC$ \\  
\hline
$\Lambda$CDM & 61.417 & 54.411& 0 & 0 \\   
$\Lambda$CDM($Rsig=0.01$) & 67.924 & 60.918 & 6.507 & 6.507 \\
$\Lambda$CDM($Rsig=0.05$) & 106.400 & 99.394 & 44.983 & 44.983 \\
$f(T)$ & 61.237 & 56.231 & -0.810 & 1.820\\
$f(T)$ ($Rsig=0.01$) & 55.791 & 50.785 & -5.626 & -3.626  \\
$f(T)$ ($Rsig=0.05$) & 53.300 & 48.294& -8.117 & -6.117  \\ 
\hline\hline
\end{tabular}
\end{center}
\caption{The information criteria BIC and AIC for the traditional NFW model in $\Lambda$CDM and effective NFW in $f(T)$ gravity, alongside the corresponding differences taking $\Lambda$CDM paradigm as reference. Here we show the results with/without the off-center effect.}
\label{tab:bic}
\end{table}

We examine the applicability of the models in the following cases: no 
off-center effect, off-center distance dispersion given by $R_{sig}=0.01$, and by $R_{sig}=0.05$. The results are presented in Table \ref{tab:bic}.
The $\Lambda$CDM scenario in the first case has AIC/BIC values of $54.411$ and $61.417$, respectively. When $R_{sig}=0.01$ the values for AIC/BIC are 
$60.918$ and $67.924$, while for $R_{sig}=0.05$  the values for AIC/BIC are 
$106.400$ and $99.394$, respectively. These results show that the standard 
$\Lambda$CDM model has the lowest BIC/AIC value without considering the 
off-center effect, and thus it prefers no off-center cases. However, when off-center is taken into account, the AIC/BIC results are all in favor of the model with $f(T)$ gravity correction. In particular, the AIC/BIC values of $f(T)$ case are $55.791$ and $50.785$ respectively when $R_{sig}=0.01$, and $53.300$ and $48.294$ respectively when $R_{sig}=0.05$. Nevertheless, we should mention here that the data used in the above analysis do not significantly constrain $R_{sig}$, and thus the optimality of $f(T)$ gravity needs to be carefully studied. However, the fact that $\Delta$AIC and $\Delta$BIC are negative, indicates that $f(T)$ is favored comparing to the $\Lambda$CDM scenario, when we additionally incorporate the off-center effects. This is one of the main result of the this work.

\section{Conclusions} \label{conclution}

In this work we used weak gravitational lensing data to test General 
Relativity and $f(T)$ gravity at galaxies scales. We considered an exact spherically symmetric solution of $f(T)$ theory under the complex tetrad. which is obtained from an approximate quadratic correction to teleparallel equivalent of General Relativity, and thus it is expected to hold for every realistic deviation from General Relativity. Firstly, following the post-Newtonian approximation, we calculated the deflection angle and the shear signal of the weak lensing under $f(T)$ gravity. In particular, quantifying the deviation from General Relativity by a single parameter $Q$, we obtained the corresponding deviation in the  gravitational potential, in the shear component and in the effective ESD profile, which is mainly affected at small scales.

We divided each stellar mass sample into blue and red to test the model's 
dependence on galaxy color. We modeled the ESD profiles using the NFW  profiles, and we found that except that the ESD profiles differ significantly in red and blue galaxies, the modified gravitational model does not depend significantly on the color of the galaxies. In addition, based on the group catalog of SDSS DR7, we used the weak lens 
data to give the tight constraints on the parameter $Q$. In the end, we 
extracted the estimation for the deviation parameter from General Relativity with the latest measurement as $Q=-2.138^{+0.952}_{-0.516}\times 10^{-5}$ $Mpc^{-2}$ at $1\sigma$ confidence. Such a constraint suggest that the deviation parameter is constrained to negative values, and the value zero is excluded at 1$\sigma$ confidence level. Therefore, $f(T)$ corrections on top of General Relativity are favoured, at least at galaxy scales.

In order to compare the fitting accuracy of $f(T)$ gravity with that of 
$\Lambda$CDM cosmology, and examine the overall efficiency of the model,  we applied information criteria, and we calculated the AIC and BIC values in three different cases. Although the dependence of $f(T)$ gravity on the off-center effect implies that its optimality needs to be carefully studied, our analysis showed that $f(T)$ gravity is more consistent 
with  observational data when the  off-center  effect is considered.   
In summary, the application of information criteria verifies our aforementioned result, that $f(T)$ gravity is more efficient than General Relativity in fitting the weak lensing data, and thus it offers a challenge to the latter.

We would like to comment here that since our results are extracted under the black hole metric solution with the point source approximation, one could study the influence of the extended source. Additionally, since our analysis applies at galactic scales, and on large scales current observations of filament lensing suggest distributions of matter that are hard to be explained under the General Relativity framework, one could try to construct $f(T)$ models that  could describe them more efficiently. Finally, one could try to perform the same analysis, quantifying the deviations from General Relativity using other frameworks, such as $f(R)$ and $f(Q)$ gravity. These subjects will be investigated in future works.

We close this manuscript by mentioning that the scale-dependent matter distribution and galaxy clustering is a powerful probe of cosmological models and gravitational theories. For instance, the Bullet Cluster \citep{Wik:2014boa}, as the first clear example of a merging galaxy cluster, provides strong evidence for the existence of dark matter by featuring at cluster scales \citep{Brownstein:2007sr}. 
The data samples used in this work mainly reflect the features of galaxy scales, which cannot effectively reflect the    cluster scales.
Nevertheless, the investigation of galaxy mergers in clusters themselves is both necessary and interesting and could shed light on the galaxy evolution and dynamics within the corresponding gravity theories, and will be performed in a future project.

\section{Acknowledgments}
We thank Geyu Mo, Zhaoting Chen, Shurui Lin, Hongsheng Zhao, Dongdong Zhang for valuable discussions. This work is supported in part by National Key R\&D Program of China (2021YFC2203100), by NSFC (12261131497, 12003029), by CAS young interdisciplinary innovation team (JCTD-2022-20), by 111 Project for ``Observational and Theoretical Research on Dark Matter and Dark Energy'' (B23042), by Fundamental Research Funds for Central Universities, by CSC Innovation Talent Funds, by USTC Fellowship for International Cooperation, by USTC Research Funds of the Double First-Class Initiative. ENS acknowledges the contribution of the LISA CosWG and the COST Actions CA18108 “Quantum Gravity Phenomenology in the multi-messenger approach'' and CA21136 “Addressing observational tensions in cosmology with systematics and fundamental physics (CosmoVerse)''. 
We acknowledge the use of computing clusters {\it LINDA}  \& {\it JUDY} of the particle cosmology group at USTC.

\software{astropy \citep{2013A&A...558A..33A,2018AJ....156..123A}}

\bibliography{ftlensing}{}
\bibliographystyle{aasjournal}



\end{document}